\documentclass[twocolumn,showpacs,amsmath,amssymb,prd,nofootinbib]
{revtex4}
\usepackage{graphicx}
\def\sss{\scriptscriptstyle}
\def\^#1{^{\sss #1}}
\def\_#1{_{\sss #1}}
\def\beq{\begin{equation}}
\def\eeqno#1{\label{#1}\end{equation}}

\def\rarrow{\rightarrow }

\def\dleft{\rlap{{\it D}}\raise 8pt\hbox{$\scriptscriptstyle\Leftarrow$}}
\def\dright{\rlap{{\it
D}}\raise 8pt\hbox{$\scriptscriptstyle\Rightarrow$}}

\def\kms{{\rm ~km~s^{-1}}}
\def\cmss{{\rm cm~s^{-2}}}
\def\kpc{{\rm ~kpc}}
\def\Mpc{{\rm ~Mpc}}

\def\msun{M\_{\odot}}
\def\az{a\_{0}}

\def\lsun{~L\_{\odot}}
\def\mlsun{~(M/L)\_\odot}
\def\l0{\ell\_{0}}

\def\s{\sigma}

\def\l{\lambda}

\def\f{\phi}

\def\xlimin{{x\rarrow\infty \atop{\raise 1pt\hbox to 30pt{\rightarrowfill}}}}
\def\limlim#1#2{{#1\rarrow #2 \atop{\raise 1pt\hbox to 30pt{\rightarrowfill}}}}

\def\RM{r\_M}

\def\RM{r\_M}
\def\h72{h\_{72}}

\def\Upb{\Upsilon\^B}
\def\Upbr{\Upb\_{r}}
\def\Upbrt{\Upb\_{r'}}

\begin{document}

\title{Testing the MOND Paradigm of Modified Dynamics with Galaxy-Galaxy Gravitational Lensing}
\author{Mordehai Milgrom} \affiliation{Department of Particle Physics and Astrophysics, Weizmann Institute of
Science, Rehovot 76100, Israel}

\begin{abstract}
MOND predicts that the asymptotic gravitational potential of an isolated, bounded (baryonic) mass, $M$, is  $\f(r)=(MG\az)^{1/2}{\rm ln}(r)$; $\az$ is the MOND acceleration constant. Relativistic MOND theories predict that the lensing effects of $M$ are dictated by $\f(r)$ as general-relativity lensing is dictated by the Newtonian potential.
Thus, MOND predicts that the asymptotic Newtonian potential deduced from galaxy-galaxy gravitational lensing
will have: (1) a logarithmic $r$ dependence, and (2) a normalization (parametrized standardly as $2\s^2$) that depends only on $M$: $\s=(MG\az/4)^{1/4}$. I compare these predictions with recent results of galaxy-galaxy lensing, and find agreement on all counts. For the ``blue''-lenses subsample (``spiral'' galaxies) MOND reproduces the observations well with an $r'$-band $M/L_{r'}\sim (1-3)\mlsun$, and for ``red'' lenses (``elliptical'' galaxies) with $M/L_{r'}\sim (3-6)\mlsun$, both consistent with baryons only. In contradistinction, Newtonian analysis requires, typically, $M/L_{r'}\sim 130\mlsun$, bespeaking a mass discrepancy of a factor $\sim 40$. Compared with the staple, rotation-curve tests, MOND is here tested in a wider population of galaxies, through a different phenomenon, using relativistic test objects, and is probed to several-times-lower accelerations--as low as a few percent of $\az$.
\end{abstract}

\maketitle

\section{introduction}
MOND \cite{milgrom83} is a theoretical framework positing strong departures from Newtonian dynamics and general relativity (GR) at low accelerations. It aims to account for the mass discrepancies in the Universe (including that associated with ``dark energy'') without invoking new entities, such as ``dark matter'' (DM). MOND introduces a new constant, $\az$, with the dimensions of acceleration, below which dynamics depart from standard dynamics: the lower the acceleration the larger the predicted discrepancy.  Reference \cite{fm12} is recent review of MOND.
\par
MOND has been amply tested in disk galaxies of all types, using rotation curves and the mass-rotational-speed relation, over an acceleration range of $\sim (0.1-10)\az$, as well as in very few elliptical galaxies (see, e.g., Ref. \cite{milgrom12}). It has also been tested in diverse pressure-supported, low-acceleration systems, such as dwarf-spheroidal satellites of the Milky Way (e.g., \cite{angus08}) and of Andromeda \cite{mm13}, tidal dwarfs \cite{gentile07,milgrom07}, and small galaxy groups \cite{milgrom02}. In galaxy clusters, MOND does reduce the mass discrepancy from a factor of $\sim 10$ to a factor of $\sim 2$. This lingering, much reduced, but systematically present discrepancy lends itself to various explanations (such as being due to yet-undetected baryons, or to neutrinos). However, until a concrete explanation is confirmed, this residual discrepancy remains a challenge for MOND (see Ref. \cite{fm12} and references therein).
\par
Some still view MOND as a mere ``phenomenological scheme'' that accounts well for observed galaxy dynamics. Partly in light of the successes of $\Lambda$CDM on large scales, many hope to see the successful MOND predictions on smaller scales explained, one day, within the DM paradigm via some, yet mysterious, connections between baryons and DM. I quite disagree with these views. First, MOND is backed by full-fledged theories on par with Newtonian dynamics in the nonrelativistic (NR) regime, and with GR in the relativistic regime, and it is not more of a ``scheme'' and less of a ``theory'' than these are. Second, as has been amply explained, the extent,  accuracy, and successes of the MOND predictions are serious challenges for the DM paradigm, even more so than its many direct conflicts with observations on small scales (see, e.g., \cite{milgrom11,fm12,kpm12} for details). It remains to be seen which of the two paradigms will prevail.
\par
The technique of galaxy-galaxy lensing (GGL) uses the statistically averaged, small distortions (weak lensing) of background-galaxy images, produced by gravitational lensing due to foreground galaxies (also averaged over large subsamples), to measure the gravitational fields around the latter (for a review see, e.g., Ref. \cite{hj08}). This method of mapping gravitational fields
is less accurate than rotation-curve analysis, and is only statistical in nature, dealing, as it does, with average properties of large samples of galaxies, not with individual ones. Yet, it offers important advantages and extends MOND testing (and probing of DM for those who think it is responsible for the mass discrepancies) in areas not accessible to other methods.
(1): In individual elliptical galaxies, strong gravitational lensing of quasars can only test MOND at small radii where accelerations are of order $\az$, and so MOND effects are small (see, e.g., Ref. \cite{sl08}). It is hardly possible to test the predictions of MOND in the low-acceleration regime--where it matters most--for reasons explained in detail in Refs. \cite{fm12,milgrom12}. Two rare exceptions are described in Ref. \cite{milgrom12}. This leaves GGL as the only method to test MOND in the very-low-acceleration (large radius) regimes of many ellipticals, albeit in a statistical manner.
(2) For disc galaxies, rotation-curve analysis affords the most powerful and accurate tests of MOND: It tests predictions of the detailed shape and of the magnitude of the accurately determined rotation curves of {\it individual} galaxies, from the baryon mass distribution alone. But, rotation curves probe discs only to intermediate radii--up to tens of $\kpc$ in some galaxies--and down to accelerations only as low as $\sim \az/10$. GGL, while cruder, extends the tests of MOND in disc galaxies, to radii several times larger, and accelerations several times lower. (MOND is seen in action at even larger distances, but similar accelerations, in the history of the Milky-Way-Andromeda system \cite{zhao13}.)
(3) GGL tests MOND by a very different technique, using unbound photons as probes, instead of bound massive particles in other techniques, and so extends the compass of MOND application and testing. (4) Unlike other methods, GGL involves aspects of  relativistic MOND.
\par
The recent results of Ref. \cite{brim13}, which map the ``DM halos'' of galaxies using GGL, provide a vary useful and ready data set. In particular, unlike most other analyses, it analyzes the data in terms of ``isothermal sphere halos'', which lends them to direct comparison with the predictions of MOND. I use these here to test MOND in the wide range of galactic radii characterizing the asymptotic, but ``isolated'' regime.
\par
An earlier MOND analysis of GGL is described in Ref. \cite{tian09}, where some tension between MOND predictions and the observations was claimed, in that in the two highest-luminosity bins galaxies required too high baryonic $M/L$ values in MOND. However, the results of Ref. \cite{brim13}, used here, are based on data rather superior to those used in Ref. \cite{tian09}, whose analysis was, in addition, beset by other issues, as discussed in Ref. \cite{fm12}.
\par
In section \ref{predictions}, I derive the MOND predictions and describe their underlying assumptions. These predictions are compared with the data in section \ref{analysis}. Section \ref{discussion} is a discussion.

\section{\label{predictions}The MOND predictions}
The MOND predictions we are testing here concern the light bending effects of a mass (a galaxy in our case) in the asymptotic and isolated regime of radii. The asymptotic regime is defined, in the MOND context, by two requirements: (1) The radii probed are beyond the region containing most of the baryonic mass. This make the predictions oblivious to details of the mass distribution, which can then be taken as a point mass $M$;\footnote{This then also obviates effects of departures from the thin-lens approximation, which occur in MOND.} (2) $r\gg \RM\equiv (MG/\az)^{1/2}\approx 11 (M/10^{11}\msun)^{1/2}\kpc$, where $\RM$ is the MOND radius of the mass ($\az=1.2\times 10^{-8}\cmss$ is used throughout). This ensures that we are deep in the MOND regime, where we can make universal predictions that are independent, for example, of the MOND interpolating function.
\par
By ``isolated'' we mean that the radii are small enough that the acceleration field is dominated by the central mass, and is not materially affected by the MOND external-field effect (EFE) from external masses, such as neighboring galaxies, clusters, or other large scale structures (see Ref. \cite{fm12} for details).\footnote{There is no detailed treatment of the EFE on lensing of photons, which traverse large distances within the mother system that produces the EFE, as well as near the affected mass itself (but see Ref. \cite{wu08}). However, since GGL measures derivatives of the potential, I assume that the same criterion that applies to bound, massive test particles applies here: The EFE is important only if the external-field acceleration is of order or larger than that of the internal field. In any event, the EFE is ignored here, and this criterion is only used for justifying this. When an EFE is present it can also mimic ellipticity in the deduce fictitious ``halo''.}
\par
We parametrize the external acceleration strength as $\eta\az$. For example, the typical field of the large-scale structure, at a random position, is estimated to have $\eta$ of a few percents.\footnote{Indicated, e.g., by the typical peculiar velocities of galaxies of a few hundred $\kms$, which would be reached, at such accelerations, in about half the Hubble time. For a more detailed estimates of $\eta$ see, e.g., Refs. \cite{milgrom98,ag01,wu08}.} We are in the isolated regime for radii within
\beq r\_{\rm isol}=\eta^{-1}\RM\approx 275 \left(\frac{M}{10^{11}\msun}\right)^{1/2}\left(\frac{\eta}{0.04}\right)^{-1}\kpc,  \eeqno{x}
which gives us a large range of radii where all conditions are satisfied.\footnote{When dealing with lensing, another requirement is of isolation from line-of-sight neighbors, which can undermine the single-lens assumption used in the analysis. This is addressed in Ref. \cite{brim13}.} At $r\_{\rm isol}$, the correction due to the EFE is of order unity.
\par
For such a regime, all existing NR, modified-gravity MOND theories, such as the nonlinear Poisson version \cite{bm84}, and QUMOND \cite{milgrom10}, predict a gravitational potential
\beq \f(r)=(MG\az)^{1/2}{\rm ln}(r).  \eeqno{i}
\par
Furthermore, the above NR MOND theories are the limits of relativistic MOND theories--such as TeVeS \cite{bek04}, MONDified Einstein aether theories \cite{zlosnik07}, BIMOND \cite{milgrom09}, and nonlocal versions \cite{deffayet11}. These theories all predict that the above NR MOND potential determines gravitational lensing in the same way as the Newtonian potential does in GR.
In other words, the procedure applied in Ref. \cite{brim13}, and other lensing analyses, to deduce the Newtonian gravitational potential (of baryons+DM) also gives the predicted MOND potential, produced by baryons alone.
\par
In the asymptotic regime, the MOND potential of Eq.(\ref{i}) strongly dominates the Newtonian potential of the baryons. Thus in a Newtonian analysis, such as in Ref. \cite{brim13}, it would be attributed to a DM halo of accumulated mass within radius $r$
\beq M_h(<r)=M\frac{r}{\RM}=\left(\frac{M\az}{G}\right)^{1/2}r. \eeqno{ii}
This is the mass distribution of a so-called singular isothermal sphere (SIS), which is one sort of halo that is fitted to the data in Ref. \cite{brim13}. The normalization is parametrized there by a SIS velocity dispersion, $\s$, such that
\beq M_h(<r)=\left(\frac{2\s^2}{G}\right)r. \eeqno{v}
Comparing Eqs. (\ref{ii}) and (\ref{v}), we see that MOND predicts that the $\s$ value deduced from the lensing data should depend only on $M$ via:
\beq \s=\left(\frac{M G\az}{4}\right)^{1/4}.\eeqno{iii}
Note that, in MOND, $\s$--used here for ease of comparison with the findings of Ref. \cite{brim13}--is not the velocity dispersion of any mass component. In particular, it is not to be confused with the baryonic velocity dispersion, $\s\_{b}$, in the galaxy. It is simply a proxy for the asymptotic (predicted constant) rotational velocity
around $M$: $2^{1/2}\s\equiv V_{\infty}=(MG\az)^{1/4}$. Thus the MOND predictions tested here are the logarithmic potential, and the $M-V_{\infty}$ relation (aka, the baryonic Tully-Fisher relation).\footnote{And, thus, we are not testing, for example, some version of the Faber-Jackson, $M-\s\_{b}$ relation, which is of a different nature. For this, MOND predicts an approximate correlation $\s\_{b}\sim (MG\az/20)^{1/4}$. So $\s\sim 1.5\s\_{b}$ is predicted.}
\par
To be directly comparable with the data of Ref. \cite{brim13}, this prediction is written in terms of the luminosity:
\beq \s=166.7~ L_{11}^{1/4}
(\Upb)^{1/4}\h72\^{-1/2}\kms.\eeqno{iv}
Here, $L_{11}\equiv L/(10^{11}h^{-2}\lsun)$, $\Upb$ is the baryonic mass-to-light ratio ($M/L$) of the galaxy in solar units, in the same photometric band where $L$ is measured, $h\equiv H_0/(100 \kms\Mpc^{-1})$, and $\h72\equiv h/0.72$ ($H_0$ is the Hubble constant).  I use these normalizations because Ref. \cite{brim13} uses everywhere $L$ (in the photometric $r'$ band) in units of $h^{-2}\lsun$, and adopt $h=0.72$.
\section{\label{analysis}Comparison with the data}
\begin{figure}
\begin{center}
\includegraphics[width=0.95\columnwidth]{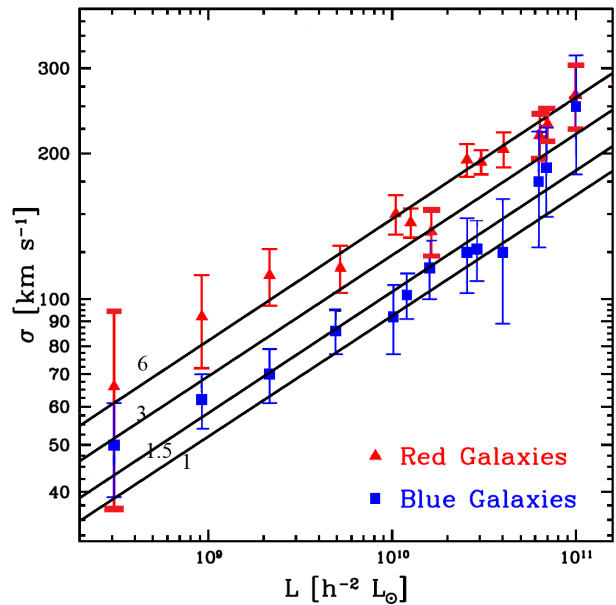}
\caption{The MOND predictions of the GGL, $\s-L_{r'}$ relations, Eq.(\ref{iv}), for baryonic mass-to-light ratios $\Upbrt=1,1.5,3,6$ (lines marked with $\Upbrt$). The measurements are reproduced from Fig. 28 of Ref. \cite{brim13}: ``blue'' lenses (blue squares), ``red'' lenses (red triangles, thick error caps where error bars overlap). The predicted lines for $\Upbrt$ of 1.5 and 6 are practically the same as the best-fit relations found in Ref. \cite{brim13} for the ``blue'' and ``red'' lenses, respectively.}\label{fig1}
\end{center}
\end{figure}
Reference \cite{brim13} used imaging data of galaxies
from the Canada-France-Hawaii Telescope Legacy WIDE survey, with good measures of the photometric redshift of the galaxies. They conducted a GGL analysis of their source galaxies by their lens galaxies, and fitted their signals as being due to three possible types of DM halos. The type relevant to us here is of the SIS halo models. For different lens-luminosity bins, Ref. \cite{brim13} fitted for the $\s$ parameter of the SIS.
The relevant lensing signal appears to come from projected radii $R\gtrsim 50\h72^{-1}\kpc$, which, comparing with the values of $\RM$, ensures, by and large, that we are in the asymptotic MOND regime.
Also, only the lensing signal within $R<140\h72^{-1}\kpc$ (projected on the sky) was used in the SIS fits. This is done to ensure relative freedom from distortion of the signal by masses other than the galaxy under study (see detailed discussion of this point in Ref. \cite{brim13}). We see from Eq.(\ref{x}), which refers to the 3-D radius, that this, by and large, also ensures isolation in the MOND sense if the external field is not too high (as would be the case, e.g., near a rich galaxy cluster). It does ensure, for example, for most galaxy positions, freedom from the EFE by large scale structure. It also justifies neglecting the EFE in groups that are not too compact and rich; for example for groups containing less than 25 galaxies of the type under study within a radius of $\sim 700\kpc$, or 100 galaxies within $\sim 1.5\Mpc$. Reference \cite{brim13} does not give enough information to assess the exact importance of the EFE, but we see from the above that, statistically, it is not expected to be important.\footnote{The fact that the SIS fits work well (see below), in itself, lends support to the unimportance of the EFE in the analysis.}
\par
The analysis is done, and results are shown, separately for the subsample of lenses classified as ``blue''--thought to be dominated by late-type, spiral galaxies--and the subsample of ``red'' galaxies--thought to be dominated by early-type, ellipticals. This segregation is particularly important when testing MOND: We are not interested in a mere phenomenological dependence of the lensing signal on luminosity; we need the dependence on the baryonic mass; so we need to convert luminosities to baryonic masses assuming reasonable mass-to-light ratios. Since galaxies in the two subsamples are known to have different $M/L$ values, such a separation is imperative.
\par
Now to the comparison with the MOND predictions. Reference \cite{brim13} makes the general statement: ``the measured gravitational shear signal is isothermal for $R \le 280 \h72\^{-1}\kpc$.'' This is consistent with the MOND prediction regarding the $r$ dependence of the lensing signal [Eqs. (\ref{i}) and (\ref{ii})].\footnote{The signal can also be fitted with other halo density laws, such as NFW, which are hard to distinguish from SIS with the data in the $R$ range used in the fit.}
For each of their bins of luminosity, $L$, in each of the two lens subsamples, the normalization of the SIS potential is best fitted for $\s$. A plot of $\s$ vs $L$ is shown in Fig. 28 of Ref. \cite{brim13}, reproduced here in Fig. \ref{fig1}.
\par
I also show in Fig. \ref{fig1} the $\s-L$ relations predicted by MOND [in Eq. (\ref{iv})], for four values of the $r'$-band $\Upbrt=1,1.5,3,6$.
\par
The ``baryonic'' $M/L$ values, which need to be used, can be larger than stellar $M/L$ values, standardly discussed in the literature, and calculated from population-synthesis models, as they must reckon with the mass in gas as well as that in stars. The difference may be quite significant in neutral-gas-reach galaxies, which would belong typically, in the low-luminosity, ``blue'' type, where $\Upb$ can be several times larger than the stellar value (see, e.g., data on this in Table 4 of Ref. \cite{ms13}). And, hot gas in ellipticals also increases $\Upb$ over the stellar value.
\par
We see that the data agree with the MOND predictions for $\Upbrt\sim 1-3$ for the ``blue'' galaxies, and $\Upbrt\sim 3-6$ for the ``red'' ones, and that within each group $\s\propto M^{1/4}$ is approximately satisfied with these $M/L$ values. In fact, Ref. \cite{brim13} gives the best fit $\s-L$ power-law correlation: for the ``blue'' galaxies: $\s\propto L^{0.23\pm 0.03}$, and for the ``red'' ones: $\s\propto L^{0.24\pm 0.03}$. Both are in very good agreement with the MOND prediction $\s\propto M^{0.25}$, if $\Upbrt$ does not vary much, systematically, within each subsample. The normalization of these fits are pinned in Ref. \cite{brim13} by $\s=\s\^*$ at $L_{11}=L^*_{11}=0.16$: $\s\^*=115\pm 3\kms$ for the ``blue'', and $\s\^*=162\pm 2\kms$ for the ``red'' subsample. MOND predicts these normalizations for $\Upbrt$ of 1.5 and 5.9, respectively. Thus, the MOND predictions shown for $\Upbrt$ of 1.5 and 6 practically coincide with the best fits of Ref. \cite{brim13}, and agree with all the data within the quoted errors.
\par
Some systematic variation of $\Upbrt$ with $L_{r'}$, within each sample (as well as scatter for a given $L_{r'}$) are consistent (but not required) by the results of Ref. \cite{brim13}. Since, by selection, the higher-redshift lenses tend to be more luminous, luminosity evolution with redshift can contribute to such systematics. Another influence could be the change in relative contribution of the gas to the baryonic mass. However, it is difficult to pinpoint the exact causes with the scant information we have. In any event, these $M/L$ variations are very minor compared with the differences between the dynamical $M/L$ values required by Newtonian/GR dynamics and by MOND, which constitutes our main result here.
\par
The $M/L$ values MOND requires are very reasonable baryonic values. For example, Ref. \cite{lintott06} shows (in its Fig. 4) {\it stellar} $M/L_r$ values for ellipticals, both measured and calculated from population synthesis, which agree well with a range of $\Upbrt\sim 3-6$ MOND requires here (they used $h=0.7$). Another pertinent recent determination of baryonic $M/L_r$ for the inner parts of compact, early-type galaxies, is Ref. \cite{conroy13}, who finds (its Fig. 3) $\Upbr$ to correlate with $\s\_{b}$, and vary between $\Upbr\sim 3$ for $\s\_{b}=100\kms$ ($\s\sim 150\kms$) and $\Upbr\sim 6$ for $\s\_{b}=250\kms$ ($\s\sim 375\kms$) (using $h=0.7$). This is quite consistent with what we see in Fig. \ref{fig1}.
MOND is thus consistent with no mass discrepancy.
In comparison, the Newtonian/GR analysis in Ref. \cite{brim13}, using halo models of finite mass gives much larger dynamical M/L values. For example, truncated isothermal spheres of $L=L^*$ are found to have a joint-sample mean $M_{dyn}/L\approx 130\h72\mlsun$, corresponding to a mass discrepancy $\gtrsim 30$.
\par
Reference \cite{brim13} also gives the best-fitted truncation radius for a truncated-isothermal-sphere halo using the combined ``blue''-``red'' sample, for the reference luminosity $L^*=3.1\times 10^{10}\h72^{-2}\lsun$: $r\_{\rm trunc}\approx 255 \h72\^{-1}\kpc$.  Taking $\Upbrt=3$ gives a MOND acceleration at this radius of $\approx 4\times 10^{-2}\az$. This is consistent with this truncation being  due to the EFE of a background field of a few percents of $\az$ [see Eq.(\ref{x})].
\par
Reference \cite{brim13} also show their $\s-L$ relations after correcting for some modeled evolution of $L$ with redshift. The changes seem rather small to make a difference in the present context. The quoted slopes become $0.26\pm 0.03$ (red) and $0.25\pm 0.03$ (blue).

\section{\label{discussion}Discussion}
We found that MOND's predictions of existing formulations match the GGL measurements and analysis of Ref. \cite{brim13}. These predictions are clear-cut and do not involve details of the theory, such as knowledge the MOND interpolating function (apart from possible influences of the EFE, which I estimated not to be important, by and large). While not as accurate and detailed as rotation-curve tests in individual disc galaxies, and while only statistical, this MOND test is a major advance: It probes, in {\it all} galaxy types, unprecedentedly large radii and very low accelerations--arguably as low as can be tested without running into the omnipresent EFE from large-scale structure.
Our results add to all the cases where MOND is shown to work well, which teach us, collectively, that baryons alone determine the whole dynamics of galaxies through the simple MOND prescription. This is quite contrary to the expectations in the DM paradigm, where the purported DM halo by far dominates the dynamics over baryons, and where the amount and distribution of baryons in the putative DM halo are determined by haphazard, violent and unpredictable processes (such as supernovae and active galactic nuclei causing losses of most of the baryons in a galaxy). How can then the puny baryons, constituting only a few percent of the total required dynamical mass, and occupying only a minute fraction of the studied volume, determine all the effects attributed to a much more massive, and much more extended DM halo, through a simple, universal relation?

\clearpage

\begin{thebibliography}{}
\bibitem{milgrom83}M. Milgrom, Astrophys. J. 270, 365 (1983).
\bibitem{fm12}B. Famaey and S. McGaugh, Living Reviews in Relativity 15, 10 (2012).
\bibitem{milgrom12}M. Milgrom,  Phys. Rev. Lett. 109, 131101 (2012).
\bibitem{angus08}G. W. Angus, Non. Not. R. Astron. Soc. 387, 1481 (2008).
\bibitem{mm13}S. McGaugh and M. Milgrom, Astrophys. J. 766, 22 (2013).
\bibitem{gentile07}G. Gentile, B. Famaey, F. Combes, P. Kroupa, H. S. Zhao, and O. Tiret,  Astron. Astrophys. Lett. 472, L25 (2007).
\bibitem{milgrom07}M. Milgrom, Astrophys. J. Lett. 667, L45 (2007).
\bibitem{milgrom02}M. Milgrom, Astrophys. J. Lett. 577, L75 (2002).
\bibitem{milgrom11}M. Milgrom, Proceedings of Science PoS(HRMS)033, arXiv:1101.5122 (2010).
\bibitem{kpm12}P. Kroupa, M. Pawlowski, and M. Milgrom, Int. J. Mod. Phys. D21, 1230003 (2012).
\bibitem{hj08}H. Hoekstra and B. Jain, Ann. Rev. Nuc. Part. Sc. 58, 99 (2008).
\bibitem{sl08}R.H. Sanders and D.D. Land, Mon. Not. R. Astron. Soc. 389, 701  (2008).
\bibitem{zhao13}H.S. Zhao, B. Famaey, F. L\"{u}ghausen, and P. Kroupa, arXiv:1306.6628 (2013).
\bibitem[Brimioulle et al. (2013)]{brim13}F. Brimioulle, S. Seitz, M. Lerchster, R. Bender, and
  J. Snigula,  Mon. Not. R. Astron. Soc. 432, 1046 (2013).
    \newpage
\bibitem{tian09}L. Tian, H. Hoekstra, and H.S. Zhao, Non. Not. R. Astron. Soc. 393, 885 (2009).
\bibitem{wu08}X. Wu, B. Famaey, G. Gentile, H. Perets, and H.S. Zhao, Mon. Not. R. Astron. Soc. 386, 2199 (2008).
\bibitem{milgrom98}M. Milgrom, Astrophys. J. Lett. 496, L89 (1998).
\bibitem{ag01}A. Aguirre, J. Schaye, and E. Quataert, Astrophys. J. 561, 550 (2001).
\bibitem{bm84}J. Bekenstein and M. Milgrom, Astrophys. J. 286, 7 (1984).
\bibitem{milgrom10}M. Milgrom, Non. Not. R. Astron. Soc. 403, 886 (2010).
\bibitem{bek04}J.D. Bekenstein, Phys. Rev. D 70, 083509 (2004).
\bibitem{zlosnik07}T.G. Zlosnik, P.G. Ferreira, and G.D. Starkman, Phys. Rev. D 75, 044017 (2007).
\bibitem{milgrom09}M. Milgrom, Phys. Rev. D 80, 123536 (2009).
\bibitem{deffayet11}C. Deffayet, G. Esposito-Farese, and R.P. Woodard, Phys. Rev. D 84, 124054 (2011).
\bibitem{ms13}S. McGaugh and J. Schombert, arXiv:1303.0320 (2013).
\bibitem{lintott06}C. J. Lintott, I. Ferreras, and O. Lahav, Astrophys. J. 648, 826 (2006).
\bibitem{conroy13}C. Conroy, A.A. Dutton, G.J. Graves, J.T. Mendel, and P.G. van Dokkum, arXiv:1306.2316  (2013).
\end{thebibliography}
\end{document}